\documentclass[11pt]{article}

\usepackage{amsmath,amsmath,amssymb}
\usepackage{amsthm}

\usepackage{mathrsfs}
\usepackage{multirow}
\usepackage{graphicx}
\usepackage{authblk}
\usepackage{indentfirst}
\usepackage{multicol}
\usepackage{tabu}
\usepackage{tabularx}
\usepackage{url}
\usepackage{fancyhdr}
\usepackage[numbers]{natbib}
\usepackage{lineno}

\usepackage{fancyhdr}
\usepackage{colortbl}
\usepackage{caption}
\usepackage{booktabs}
\usepackage{hyperref}
\hypersetup{colorlinks=true,linkcolor=blue,anchorcolor=blue,citecolor=blue}

\usepackage[table,xcdraw]{xcolor}
\usepackage{longtable}
\usepackage{hhline}
\usepackage{alltt}
\usepackage{tgcursor}
\usepackage{xr}
\usepackage{tikz-cd}

\makeatletter
\newcommand*{\addFileDependency}[1]{
	\typeout{(#1)}
	\@addtofilelist{#1}
	\IfFileExists{#1}{}{\typeout{No file #1.}}
}
\makeatother

\begin{document}
	\setlength{\parindent}{2em}
	\title{Mayer-homology learning prediction of  protein-ligand binding affinities}

	\author[1]{Hongsong Feng}
	\author[1]{Li Shen }
	\author[2,1]{Jian Liu}
	\author[1,3,4]{Guo-Wei Wei \thanks{Corresponding author: weig@msu.edu}}
	\affil[1]{Department of Mathematics, Michigan State University, East Lansing, MI 48824, USA}
	\affil[2]{Mathematical Science Research Center, Chongqing University of Technology, Chongqing 400054, China}
	\affil[3]{Department of Biochemistry and Molecular Biology, Michigan State University, MI, 48824, USA}
	\affil[4]{Department of Electrical and Computer Engineering, Michigan State University, MI 48824, USA}
	\renewcommand*{\Affilfont}{\small\it}
	\renewcommand\Authands{ and }
	\date{}
	
	
	\maketitle
	\begin{abstract}
Artificial intelligence-assisted drug design is revolutionizing the pharmaceutical industry. Effective molecular features are crucial for accurate machine learning predictions, and advanced mathematics plays a key role in designing these features. Persistent homology theory, which equips topological invariants with persistence, provides valuable insights into molecular structures. The calculation of Betti numbers is based on differential that typically satisfy \(d^2 = 0\). Our recent work has extended this concept by employing Mayer homology with a generalized differential that satisfies \(d^N = 0\) for \(N \geq 2\), leading to the development of persistent Mayer homology (PMH) theory and richer topological information across various scales. In this study, we utilize PMH to create a novel multiscale topological vectorization  for molecular representation, offering valuable tools for descriptive and predictive analysis in molecular data and machine learning prediction. Specifically, benchmark tests on established protein-ligand datasets, including PDBbind-2007, PDBbind-2013, and PDBbind-2016, demonstrate the superior performance of our Mayer homology models in predicting protein-ligand binding affinities.
  \end{abstract}	
	
		
	\newpage
	
	\section{Introduction}
			
	Drug design and discovery is a lengthy and costly process, and  the process is plagued by a high failure rate \cite{fleming2018computer}. Developing a new drug can take a decade and cost billions of dollars before it reaches the market. However, the accumulation of extensive experimental biological data and advancements in machine learning algorithms are rapidly transforming AI-based drug design, offering the potential to revolutionize the field. For instance, determining protein–ligand binding affinity is a critical challenge in drug discovery. Machine learning models excel at capturing complex, non-linear relationships in data and provide superior accuracy in predicting binding affinity compared to traditional models \cite{li2015improving,feinberg2018potentialnet}.
	
	In AI-assisted drug design and discovery, molecular descriptors are crucial to the performance of machine learning models and are integral to various stages of the drug discovery process \cite{ballester2010machine,pan2022aa,wang2017improving,gu2023can}. They play a fundamental role in quantitative structure–activity relationship (QSAR) \cite{soares2022re} and quantitative structure–property relationship (QSPR) analyses \cite{sato2021comparing}. Two-dimensional (2D) and three-dimensional (3D) molecular descriptors are widely used for machine learning predictions. Popular 2D molecular descriptor generation approaches include substructure key-based fingerprints, topological or path-based fingerprints, circular fingerprints, pharmacophore fingerprints \cite{durant2002reoptimization}, and autoencoded fingerprints \cite{winter2019learning}. These molecular descriptors can typically be extracted from molecular simplified molecular-input line-entry system (SMILES) strings without 3D structure information, and they are primarily used to represent small molecules. However, they do not perform well for macromolecules with complex 3D structures. The 3D molecular descriptors consider molecular structures, as well as chemical, physical, and biological properties \cite{todeschini2008handbook}. 3D molecular descriptors are superior to 2D ones in scenarios where the three-dimensional structure of a molecule is a key determinant of its biological function, such as protein-ligand interactions, stereochemistry, molecular docking and virtual screening, and conformational analysis. Deep learning methods, such as autoencoders \cite{winter2019learning}, transformers \cite{chen2021extracting}, graph neural networks (GNNs) \cite{nguyen2021graphdta}, and convolutional neural networks (CNNs) \cite{wallach2015atomnet,ragoza2017protein,jimenez2018k}, have also been employed for molecular feature generation, producing highly competitive molecular features.

	The complexity of biomolecular structure, function, and dynamics often renders structural representation inconclusive, inadequate, inefficient, and, at times, intractable. These challenges necessitate innovative design strategies for representing macromolecules. Recently, advanced mathematical tools such as topological data analysis (TDA) have been employed for biomolecular characterization \cite{cang2017topologynet}. These mathematical approaches excel in abstracting and representing intrinsically complex molecular structures, making them particularly suitable for molecular featurization \cite{cang2017topologynet,bi2024persistent}. Although TDA is a relatively new field, it has rapidly emerged as a powerful methodology in data science. The essence of TDA lies in extracting significant topological invariants and geometric shapes, capturing the nuanced patterns and relationships embedded within the data. Persistent homology theory \cite{carlsson2007theory,edelsbrunner2008persistent,zomorodian2004computing,carlsson2010zigzag} is the primary workhorse in TDA. It utilizes a filtration process that gradually builds up a family of topological spaces from the data by adding simplices (basic geometric shapes like points, edges, triangles, etc.) one at a time, based on parameters such as distance or density. Multiscale representation is reflected in two key aspects: tracking the lifespan of topological invariants across scales—such as connected components, loops, and voids—and describing these invariants, including Betti numbers, which count the number of topological features at different scales. These two aspects are essentially equivalent but provide quantitative analysis from different perspectives.

	Persistent homology is particularly useful for characterizing molecular data and distinguishes itself from traditional molecular descriptors \cite{cang2017topologynet}. Its integration with machine learning algorithms has led to significant advancements in various stages of drug design such as predicting protein–ligand binding affinity \cite{cang2017topologynet} and changes in protein stability upon mutation \cite{wang2020topology}. One of the most compelling pieces of evidence for the effectiveness of persistent homology-based molecular descriptors is their significant success in the D3R Grand Challenge \cite{nguyen2019mathematical}, a worldwide computer-aided drug discovery competition. The power of TDA is further exemplified in the topological deep learning (TDL) paradigm \cite{cang2017topologynet}, where topological molecular featurization is integrated with deep neural networks. The early success of mathematical representations and TDL for biomolecular data was documented in a review  \cite{nguyen2020review}. Recently, TDL has uncovered the mechanisms of SARS-CoV-2 evolution \cite{chen2020mutations}.
				 	
	Despite its tremendous success, persistent homology theory still faces some drawbacks. Persistent homology can lead to an oversimplification of geometric information from the given data, resulting in a significant loss of crucial details. For example, while it captures cavities and loops, it lacks the precision to describe their shape, depth, or orientation, potentially treating distinct geometries—such as a shallow, wide pocket and a deep, narrow one—as equivalent. To address these limitations, persistent Laplacian theories \cite{wang2020persistent,meng2021persistent,wei2023persistent} were introduced. While retaining the ability to capture topological invariants like persistent homology, persistent Laplacians also reveal additional geometric information through the non-harmonic spectrum of their matrices. Their value as molecular descriptors is demonstrated by their accurate forecasting of emerging dominant SARS-CoV-2 variants BA.4/BA.5 \cite{chen2022persistent}. Other valuable topological or geometric data analysis techniques with superb performance include persistent de Rham-Hodge Laplacians \cite{su2024persistent}, and  neighborhood path complex  \cite{liu2023neighborhood}. 

	In this study, we explore the potential of persistent Mayer homology (PMH) for characterizing molecules. Unlike usual simplicial homology, which uses a boundary operator \( d \) that satisfies \( d^2 = 0 \), Mayer Homology defined on simplicial complexes employs a boundary operator \( d \) where \( d^N = 0 \) \cite{mayer1942new}. Mayer Homology reduced to simplicial homology when \( N = 2 \). The boundary operator is essential for constructing algebraic groups used to identify properties related to connected components, rings, or cavities in simplicial homology. Mayer homology theory provides enhanced Betti number information by varying the value of \( N \), revealing more detailed topological and geometric features of a space \cite{shen2024persistent}. A multiscale topological representation from PMH is obtained through the filtration process, in which a series of nested simplicial complexs is generated. During this process, Betti numbers or the persistence of topological invariants are calculated and used to construct molecular descriptors. These topological features are then combined with a gradient boosting decision tree algorithm to build predictive models. We evaluate the performance of our PMH-based machine learning models on three well-established protein-ligand binding affinity datasets: PDBbind-v2007, PDBbind-v2013, and PDBbind-v2016 \cite{liu2015pdb}. Our models demonstrate exceptional performance, which can be further enhanced by incorporating molecular sequence-based descriptors generated using natural language processing techniques. Compared to the existing literature, our models achieve state-of-the-art performance.

	\section{Results}\label{sec:results}

	\subsection{Overview of persistent Mayer homology (PMH)}
	
	\begin{figure}[!htb]
		\centering
				\includegraphics[width=0.9\textwidth]{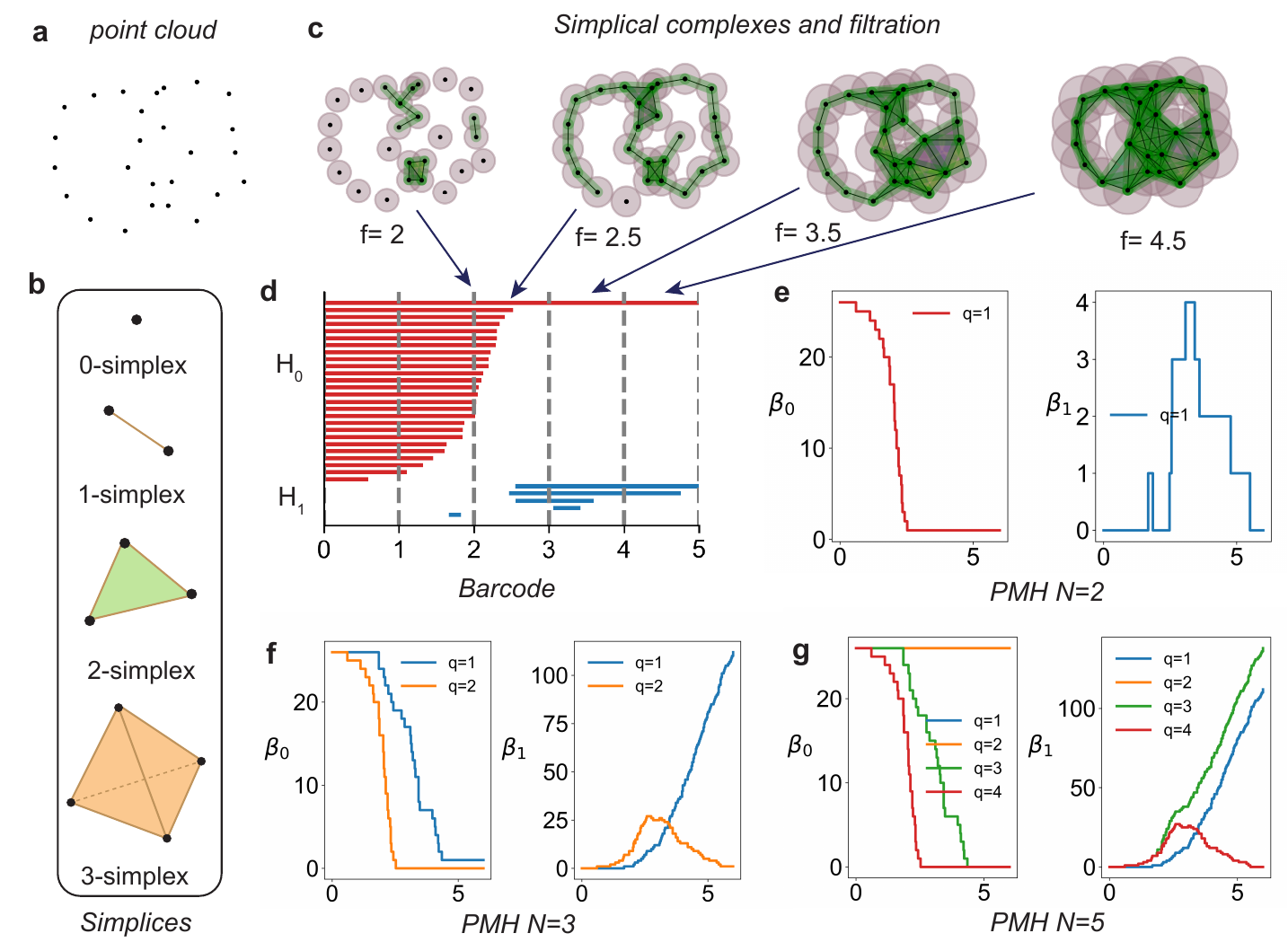}
		\caption{The persistent Mayer homology representation for a point cloud based on VR complex.
			 {\bf a}: A 2D point cloud. {\bf b}: The representation of simplices in dimension $n=0,1,2,3$. {\bf c}: A filtration of simplicial complexes obtained from the point cloud. {\bf d}: The barcode of dimension 0 and 1 corresponding to the filtration process in {\bf c}. The filtration parameter is defined to be the diameter of circles around given points. {\bf e}: The Betti numbers $\beta_0$ and $\beta_1$ calculated from persistent Mayer homology (PMH) for $N=2$. {\bf f}: The Betti numbers $\beta_0$ and $\beta_1$ calculated from persistent Mayer homology (PMH) for $N=3$. {\bf g}: The Betti numbers $\beta_0$ and $\beta_1$ calculated from persistent Mayer homology (PMH) for $N=5$. The curves at $q=2$ and $q=3$ coincides in {\bf g}.
		}
		\label{Fig:representation-PMH}
	\end{figure}

As mentioned early, Mayer homology of simplicial complex reduces to simplicial homology when $N$ is taken to 2. We will begin with a brief review of simplicial complexes, the classical homology of simplicial complexes, and then generalize the discussion to Mayer homology.

Simplicial complex is a well-known topological model in data science, with notable examples including the Vietoris-Rips complex, \v{C}ech complex, and Alpha complex. A simplicial complex is composed of a collection of simplices following specific combinatorial rules. An $n$-simplex is the convex hull formed by $n$ geometrically independent points. For example, a 0-simplex is a vertex, a 1-simplex is an edge, a 2-simplex is a triangle (with a solid interior), and a 3-simplex is a solid tetrahedron, as illustrated in \autoref{Fig:representation-PMH}b.

Let \( K = \{K_n\} \) be a simplicial complex, where \( K_n \) is the set of \( n \)-simplices. For a given field \( \mathbb{F} \), we denote by \( C_n(K) \) the \( \mathbb{F} \)-vector space generated by the \( n \)-simplices in \( K_n \). The collection of such groups \( (C_n(K))_{n \geq 0} \) gives rise to a chain complex, with the differential defined by
\[
d_n \langle v_0, v_1, \ldots, v_n \rangle = \sum_{i=0}^n (-1)^i \langle v_0, \ldots, \hat{v}_i, \ldots, v_n \rangle,
\]
where \( \langle v_0, v_1, \ldots, v_n \rangle \) is an \( n \)-simplex, and \( \hat{v}_i \) denotes the omission of the \( i \)-th vertex. It can be verified that \( d_{n-1} \circ d_n = 0 \). Thus, we obtain a chain complex
\begin{equation}
\cdots \stackrel{d_{n+2}}{\longrightarrow} C_{n+1}(K) \stackrel{d_{n+1}}{\longrightarrow} C_n(K) \stackrel{d_n}{\longrightarrow} C_{n-1}(K) \stackrel{d_{n-1}}{\longrightarrow} \cdots.
\end{equation}

We denote \( \ker d_n = \{x \in C_n(K) \mid d_n x = 0\} \) and \( \mathrm{im}~d_n = \{d_n x \mid x \in C_n(K)\} \). Since \( d_{n-1} \circ d_n = 0 \), it follows that \( \mathrm{im}~d_{n+1} \subseteq \ker d_n \). The homology of the simplicial complex \( K \) is then defined by
\begin{equation}
H_n(K) := \frac{\ker d_n}{\mathrm{im}~d_{n+1}}, \quad n \geq 0.
\end{equation}
The rank of \( H_n(K) \) is the Betti number \( \beta_n \). Betti numbers are widely used topological invariants of simplicial complexes. The geometric interpretation of \( \beta_0 \), \( \beta_1 \), and \( \beta_2 \) corresponds to the number of connected components, loops, and cavities, respectively.

The key idea of persistent homology is to introduce multi-scale information, which is provided by the filtration of simplicial complexes. For a given point cloud data set, the most common filtration of simplicial complexes is the Vietoris-Rips (VR) complex, as illustrated in \autoref{Fig:representation-PMH}c. Topological features at different scales exhibit a certain kind of persistence, meaning that homology generators at smaller scales may persist as homology generators at larger scales, thereby giving rise to persistent homology generators. The scale at which a generator is born is referred to as its birth time, while the scale at which it disappears is known as its death time. The topological features of persistent homology are represented by bars that record the birth and death times of homology generators, as shown in \autoref{Fig:representation-PMH}d, corresponding to the barcode of the filtration of simplicial complexes in \autoref{Fig:representation-PMH}d. 

Unlike classical homology theories, the Mayer homology theory explored in this study generalizes the calculation of Betti numbers by utilizing the differential \( d^N = 0 \) with an integer \( N \geq 2 \) on the \( N \)-chain complex (see \hyperref[sec:Methods]{Methods}). This approach allows us to obtain a family of homology groups \( H_{n,q}(K) \) for a simplicial complex, where \( n \) is the dimension and \( 1 \leq q \leq N-1 \) corresponds to the Mayer degree. The homology groups \( H_{n,q}(K) \) are referred to as Mayer homology. The Betti numbers associated with Mayer homology are termed the \emph{Mayer Betti numbers} of the simplicial complex, denoted by \( \beta_{k,q} \). For \( N=2 \), the Mayer degree \( q \) can only be \( q=1 \), which means that for a fixed dimension \( n \), there is only one homology group, which is consistent with the usual homology groups of a simplicial complex. For general \( N \), Mayer homology reveals more information than classical homology, offering potentially valuable geometric and topological features for applications. Beyond contributing to a unified mathematical framework for homology theory, Mayer homology and the associated Betti numbers provide valuable tools for analyzing the topological space of a given data set.

The Betti numbers for each simplicial complex are recorded in the barcode diagram shown in \autoref{Fig:representation-PMH}d. For example, the number of red lines in \autoref{Fig:representation-PMH}d at a filtration parameter of 2 corresponds to \( \beta_{0} \). The Betti number \( \beta_{n}:[0,+\infty)\to \mathbb{N} \) can be regarded as a function with the filtration parameter as its variable. Such a function is referred to as a \emph{Betti curve}. \autoref{Fig:representation-PMH}e shows the Betti curves for \( N=2 \), with the red line representing the Betti curve \( \beta_{0} \) and the blue line representing the Betti curve \( \beta_{1} \). Additionally, \autoref{Fig:representation-PMH}f and \autoref{Fig:representation-PMH}g present the Betti curves for Mayer homology with \( N=3 \) and \( N=5 \), respectively. Each plot contains multiple curves because, in the case of Mayer homology, \( \beta_{n,q} \) forms a curve for each \( 1 \leq q \leq N-1 \). It is worth noting that when \( N=5 \), the \( \beta_{k,2} \) and \( \beta_{k,3} \) align with each other as shown in \autoref{Fig:representation-PMH}f. The comparison of these figures highlights the richer topological and geometric features of Mayer Betti numbers. More detailed concepts and theorems associated with Mayer homology can be found in the \hyperref[sec:Methods]{Methods}.

\subsection{PMH-based element interactive molecular representation}

	\begin{figure}[!htb]
	\centering
	\includegraphics[width=0.9\textwidth]{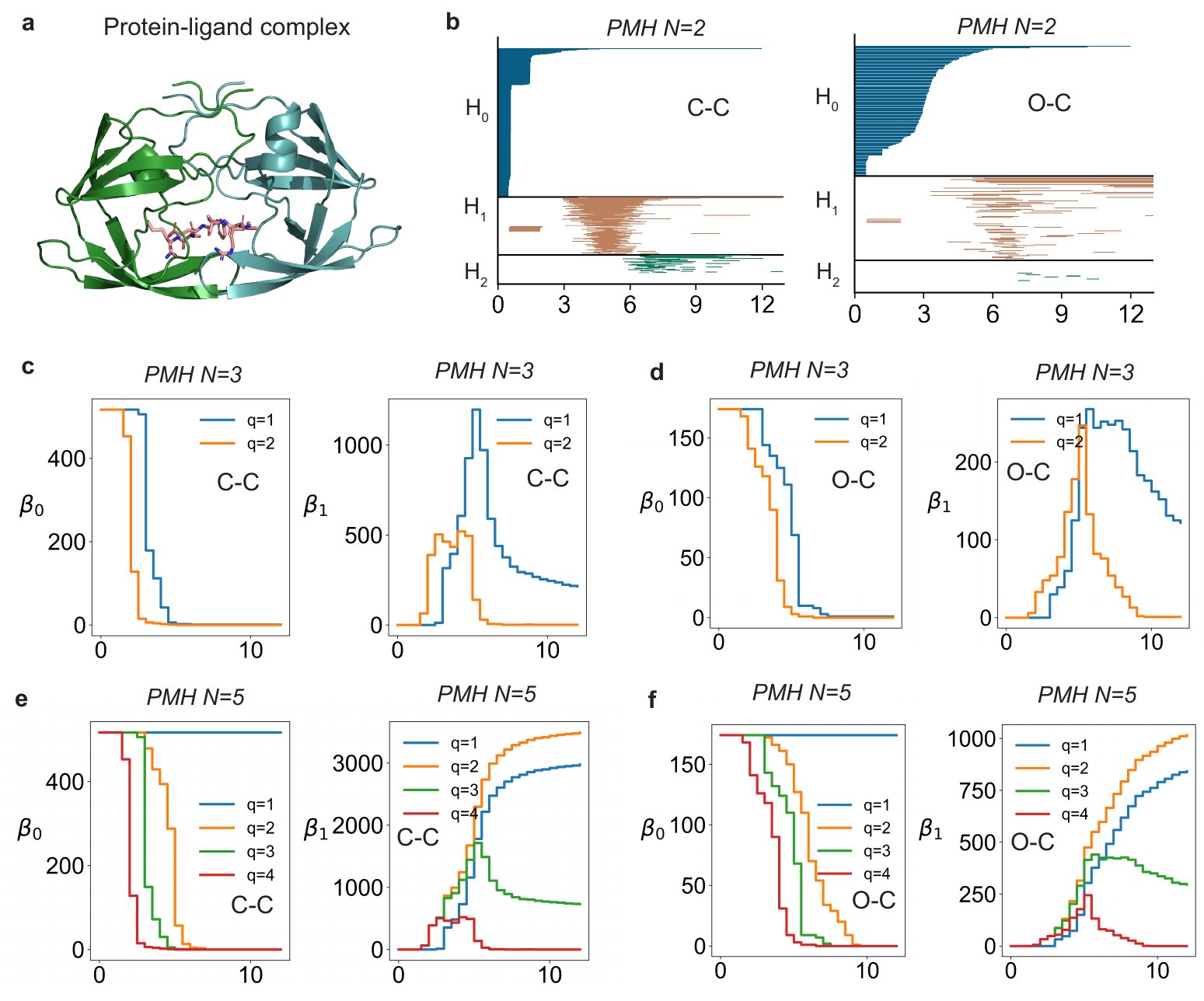}
	\caption{Persistent Mayer homology characterization for a protein-ligand complex (PDBID: 1A94) on alpha complex. {\bf a}: The 3D structure of protein 1A94. {\bf} The barcodes of different dimensions for a pair of atom sets in protein 1A94 with PMH (N=2). The first letter in C-C or O-C stands from atom group from protein and the second one indicates atom group from the ligand. {\bf c-d}: The $\beta_0^q$ and $\beta_1^q$ calculations for the atom groups in {\bf b} using  PMH with N=3 and q=1,2,$\cdots$,N-1. {\bf e-f}: The $\beta_0^q$ and $\beta_1^q$ calculations for the atom groups in {\bf b} using  PMH with N=5 and q=1,2,$\cdots$,N-1.
	}
	\label{Fig:molecular-1A94}
\end{figure}

Atomic coordinates in molecules can be viewed as point cloud data. Persistent Mayer homology is well-suited for characterizing molecular structures, and a multiscale topological representation can be obtained through a filtration process. The resulting persistent features effectively capture the hierarchical and multiscale properties of biomolecular structures and interactions. Various intramolecular and intermolecular interactions exist within molecular structures, characterized by different forces such as covalent bonds, van der Waals forces, electrostatic interactions, hydrophobic interactions, and hydrophilic interactions. To this end, we follow the element interaction characterization for pairwise atom groups \cite{cang2018integration} and use persistent Mayer homology to analyze these element-specific topological data structures. A cutoff distance of 12 \AA~is applied to extract the protein atoms around the ligand, considering that intermolecular interactions predominantly occur in the binding pocket region.

\autoref{Fig:molecular-1A94}b displays the PMH ($N=2$) barcodes for C-C and O-C atom groups in the protein-ligand complex (PDBID: 1A94), with the simplicial complex constructed using the alpha complex. The persistence and variance of the $\beta_0$, $\beta_1$, and $\beta_2$ information are revealed. The ligand has more carbon atoms than oxygen atoms, leading to the faster decay of the $\beta_0$ value during filtration for C-C atom groups. Persistent attributes associated with $\beta_1$ and $\beta_2$ are also distinguishable in the characterization of C-C and O-C atom groups. The changes in $\beta_{k,q}$ values from PMH with $N=3$ and $N=5$ for C-C groups are shown in \autoref{Fig:molecular-1A94}c and \autoref{Fig:molecular-1A94}e. The changes for O-C groups are exhibited in \autoref{Fig:molecular-1A94}d and \autoref{Fig:molecular-1A94}f. Unlike the PMH characterization for 2D point clouds, which shows overlapping curves, there are distinct $\beta_{0,q}$ or $\beta_{1,q}$ curves in \autoref{Fig:molecular-1A94}d and \autoref{Fig:molecular-1A94}f for $N=5$. These PMH ($N=3$ or $N=5$) Betti changes for these atom groups tend to plateau when the filtration parameter reaches 10 \AA, or even as early as 5 \AA. Therefore, it is sufficient to collect the Betti information with the filtration parameter ranging from 0 \AA~to 10 \AA. For PMH ($N=2$) or traditional persistent homology characterization of the protein-ligand complex, persistent attributes analysis extends to an upper filtration parameter of 12 \AA.

It is observed that the $\beta_{0,1}$ and $\beta_{0,2}$ curves in \autoref{Fig:molecular-1A94}c resemble the $\beta_{0,3}$ and $\beta_{0,4}$ curves in \autoref{Fig:molecular-1A94}e. A similar pattern is seen between \autoref{Fig:molecular-1A94}d and \autoref{Fig:molecular-1A94}f. However, there are subtle numerical differences along the filtration. The $\beta_{0,1}$ and $\beta_{0,2}$ curves, along with the distinct $\beta_{1,q}$ curves, still differentiate PMH ($N=5$) from PMH ($N=3$).

A multiscale molecular representation can be obtained either by directly using PMH Betti numbers or by extracting useful statistical information from barcodes. Persistence bars represent the persistence of topological invariants in nested simplicial complexes, from which PMH Betti numbers can be directly read. Molecular features can be designed by collecting the Betti numbers at a set of filtration parameters. However, the inconsistent number of atoms across atom groups or molecules makes barcodes not directly suitable for scalable representation learning. Various stable learning strategies for topological data analysis have been proposed, such as persistent landscapes \cite{bubenik2015statistical} and persistent images \cite{adams2017persistence}. The bin-spaced statistical functions \cite{cang2017topologynet}, incorporating the maximum, minimum, average, and standard deviation of barcodes, provide a reliable and effective vector representation. This approach offers competitive descriptive capacity and the advantage of scalable modeling. We utilize both the Betti numbers from PMH and barcodes to design molecular features.

To address computational efficiency, simplicial complexes using alpha complexes are primarily considered for PMH with \( N > 2 \). For PMH with \( N = 2 \), both VR complexes and alpha complexes can be utilized. When VR complexes are used, we incorporate physical properties in addition to the original molecular structure data to ensure that sufficient molecular interactions are captured. Technically, the filtration process and persistent Mayer homology are induced using either the Euclidean distance metric in space or a kernel function-defined correlation matrix for a group of atomic coordinates. Collectively, these methods enhance our PMH theory-based molecular representation learning. We provide more details about our PMH features in the following section.

\subsection{PMH learning models for drug design}

\begin{figure}[!htb]
	\centering
	\includegraphics[width=1\textwidth]{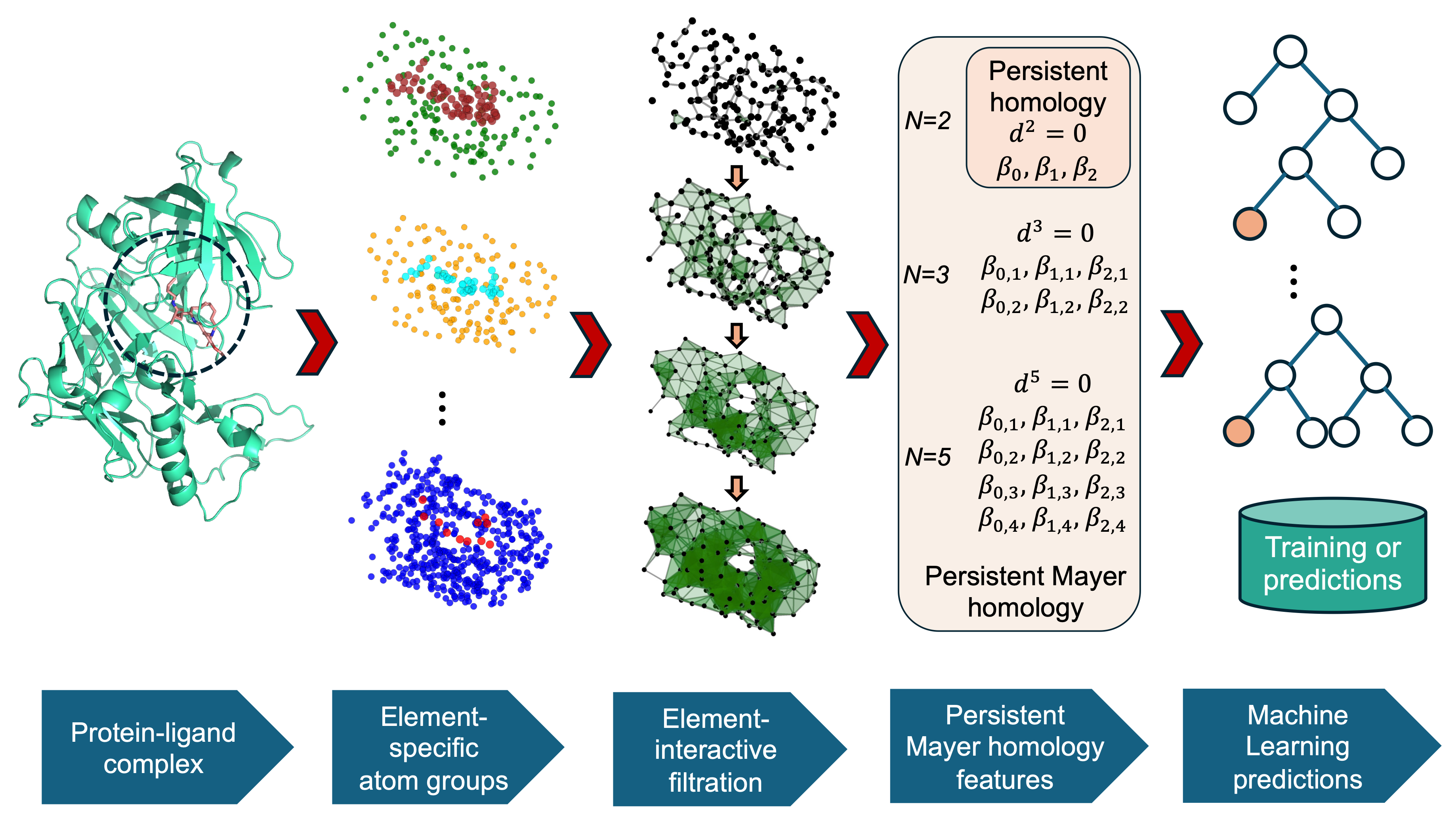}	\caption{The illustration of persistent Mayer homology feature extraction for a protein-ligand complex (PDBID: 1A94) and the subsequent machine learning model development.}
	\label{Fig:ml-concepts}
\end{figure}

\subsubsection{PMH-based multiscale molecular vectorization}

We utilize element-interactive PMH representation learning for biomolecular data, as discussed above. This strategy captures crucial biological information and enhances characterization capacity, as validated by extensive modeling work \cite{cang2017topologynet,nguyen2019agl,meng2021persistent}. Specifically, for a protein-ligand complex, the types of elements considered for proteins are \( S_P = \{ \text{C}, \text{N}, \text{O}, \text{S} \} \), and for ligands, they are \( S_L = \{ \text{C}, \text{N}, \text{O}, \text{S}, \text{P}, \text{F}, \text{Cl}, \text{Br}, \text{I} \} \). Therefore, we can have up to 36 element combinations and design interactive PMH features accordingly. The interactions between all the ligand atoms and protein atoms near the binding pocket can also be characterized by PMH.

We denote \( S_{X-Y}^c \) as the set of atoms consisting of \( X \) types of atoms in the protein and \( Y \) types of atoms in the ligand, where the distance between any pair of atoms in these two groups is within a cutoff \( c \):
\begin{align}
S^c_{X-Y} = \{a|a \in X, \min_{b\in Y} \text{dis}(a, b) \leq c\}\cup \{b|b \in Y\},
\end{align}
where a and b denote atoms. We also consider all heavy atoms in the ligand together with all heavy atoms in the protein that are within the cutoff distance \( c \) from the ligand molecule, and denote this set as \( S^c_{all} \). Similarly, we denote the set of all heavy atoms in the protein that are within the cutoff distance \( c \) from the ligand molecule as \( S^c_{pro} \).

Both the correlation matrix and the Euclidean distance matrix are used for the VR complex-induced persistent homology (PMH) (\(N=2\)). We use \(A(i)\) to indicate the affiliation of an atom with index \(i\) in a group of atoms from either the protein or the ligand. We define four types of matrices as follows.

\begin{itemize}
	\item ${\rm FRI}_{\tau,\nu}^{agst}$:
	\begin{equation}\label{eq:FRI_against}
		d(i,j) =
		\begin{cases}
			1 - e^{-(r_{ij}/\eta_{ij})^\kappa}, \: &A(i) \neq A(j) \\
			d_{\infty}, &A(i)=A(j)
		\end{cases}
	\end{equation}
	
	\item ${\rm FRI}_{\tau,\nu}$:
	\begin{equation}\label{eq:FRI}
		d(i,j) = 1 - e^{-(r_{ij}/\eta_{ij})^\kappa}
	\end{equation}	
	
	\item ${\rm EUC}^{agst}$:
	\begin{equation}\label{eq:EUC_against}
		d(i,j) =
		\begin{cases}
			r_{ij}, \: & A(i)\neq A(j) \\
			d_{\infty}, & A(i)=A(j)
		\end{cases}
	\end{equation}
	\item ${\rm EUC}$:
	\begin{equation}\label{eq:EUC}
		d(i,j) = r_{ij}.
	\end{equation}
\end{itemize}
 \autoref{eq:FRI_against} is inspired by the development of the flexibility-rigidity index (FRI) theory \cite{xia2013multiscale}, which utilizes a decaying radial basis function to effectively quantify atomic interactions. The parameter \(r_{ij}\) represents the Euclidean distance between atoms with indices \(i\) and \(j\), and \(\eta_{ij} = \tau \cdot (r_i + r_j)\), where \(k\) and \(\tau\) are positive adjustable parameters that control the decay rate of the exponential kernel, allowing us to model interactions with different strengths. Here, \(\eta_{ij}\) is the characteristic distance between the \(i\)th and \(j\)th atoms and is typically set as the sum of the van der Waals radii of the two atoms. The exponential kernel function is non-negative and strictly monotonically decreasing with respect to the Euclidean distance between a pair of atoms. When the Euclidean distance between two atoms is close to 0, their correlation distance \(d(i,j)\) approaches 1. Conversely, when the atoms are far apart, \(d(i,j)\) approaches 0. This ensures that the correlation matrix is well-defined. We use the superscript \(agst\) to distinguish correlations between atoms from the same or different affiliations. When both atoms are within the same molecule, their correlation distance is set to infinity. This approach excludes intramolecular interactions and highlights the intermolecular interactions between proteins and ligands, which are then represented in the construction of VR simplices and ultimately aid in characterizing these interactions through persistent Mayer homology (PMH).

 In contrast, the correlation matrix defined by \autoref{eq:FRI} captures both physical and chemical information from intramolecular and intermolecular interactions. Furthermore, \autoref{eq:EUC_against} and \autoref{eq:EUC}, which are based on the Euclidean distance metric, provide a better characterization of molecular 3D structures. The ${\rm EUC}^{agst}$ metric places greater emphasis on the shape derived from intermolecular 3D data and is used in conjunction with alpha complexes for our PMH analysis. We primarily use PMH(N=2) and PMH(N=5) to extract molecular features, employing five different feature extraction strategies as shown in \autoref{tb:feature}. Consequently, for each protein-ligand complex, we generate five feature vectors: the first four are derived from PMH(N=2), while the final vector is based on PMH(N=5).

\begin{table*}[t]
	\centering
	\begin{tabular}{p{0.3cm}| p{5cm} l p{10cm}}
		\hline
		I & ${\rm PMH2}(P_{\rm ep-el}^{12},{\rm FRI}^{agst},{\rm VR})$ & \multirow{2}{10cm}{Length sum of all Betti-0 bars.}  \\
		& \centering ep $\in S_P$, el $\in S_L$ &  \\ \hline
		 & ${\rm PMH2}(P_{\rm all}^6,{\rm FRI},{\rm VR})$ & \multirow{2}{10cm}{Length sum and birth sum of Betti-0, Betti-1, and Betti-2 bars for protein, complex, as well as the sum differences between protein and complex.}  \\
		II& ${\rm PMH2}(P_{\rm pro}^6,{\rm FRI},{\rm VR})$ & \\
		& & \\ \hline
		III & ${\rm PMH2}(P_{\rm ep-el}^{12},{\rm EUC}^{agst},{\rm VR})$ & \multirow{2}{10cm}{Counts of Betti-0 bars with `death' values within each interval: $[0,2.5]$, $[2.5,3]$, $[3,3.5]$, $[3.5,4.5]$, $[4.5,6]$, $[6,12]$.}  \\
		& \centering ep $\in S_P$, el $\in S_L$ &  \\ \hline
		 & \multirow{2}{*}{${\rm PMH2}(P_{\rm all}^9,{\rm EUC},{\rm Alpha})$} & \multirow{4}{10cm}{Length sum of Betti-1 and Betti-2 bars with `birth' values within each interval: $[0,2]$, $[2,3]$, $[3,4]$, $[4,5]$, $[5,6]$, $[6,9]$. The sum differences between complex and protein are also considered.}  \\
		IV& & \\
		& \multirow{2}{*}{${\rm PMH2}(P_{\rm pro}^9,{\rm EUC},{\rm Alpha})$} &  \\
		& & \\ \hline
		 & ${\rm PMH5}(P_{\rm ep-el}^{12},{\rm EUC},{\rm Alpha})$ & \\
		& \multirow{2}{*}{ep $\in S_P \setminus \{C\}$, el $\in S_L \cup \{H\}$} & \multirow{2}{10cm}{$\beta_{k,q}$ (k=0,1, q=1,2,$\cdots$,4) over filtration parameter range from 0 to 10 with stepsize of 0.2.} \\
		V& & \\
		& \multirow{2}{*}{ep $\in \{C\}$, el $\in S_L \cup \{H\}$} & \multirow{2}{10cm}{$\beta_{k,q}$ (k=0,1, q=1,2,$\cdots$,4) over filtration parameter range from 0 to 8 with stepsize of 0.5.}\\
		& & \\ \hline
	\end{tabular}
	\caption{Molecular feature extraction with PMH. PMH2 and PMH5 indicates the PMH on 2-chain and 5-chain complex, respectively. The first argument in PMH2 or PMH5 specifies the group of molecular coordinate data, while the second argument denotes the correlation or Euclidean distance matrix. The third argument indicates the type of complex used to construct simplical complex.}
	\label{tb:feature}
\end{table*}

\subsubsection{PMH  learning models for binding affinity prediction}

We demonstrate the learning capacity of the proposed PMH through protein-ligand binding affinity prediction, a critical problem in drug discovery. We consider three well-established PDBbind datasets \cite{liu2015pdb}, including PDBbind-v2007, PDBbind-v2013, and PDBbind-v2016. These datasets contain a collection of 3D structures for protein-ligand complexes and their experimental binding affinities and have been widely used to test new methods \cite{rana2023geometric,rana2022eisa,liu2022dowker}. Detailed information about the data size for the three datasets and the related training-test splits can be found in \autoref{table:datasets-information}. Based on the 3D structures, each protein-ligand complex is represented by five sets of molecular vectors according to \autoref{tb:feature}. In our implementation, feature sets I-IV are concatenated into a long vector representation, while feature set V is used as a separate vector representation. These two vectors are combined with the gradient boosting decision tree (GBDT) algorithm to build regression models, resulting in model-PMH2 and model-PMH5. The GBDT hyperparameters used for modeling are listed in \autoref{table:GBDT-parameters}. A general workflow of our PMH featurization and the resulting machine learning modeling is provided in \autoref{Fig:ml-concepts}.

The final PMH modeling prediction is determined by the consensus of the predictions from the two models. We build models twenty times with different random seeds and use two evaluation metrics: Pearson correlation coefficient (R) and root mean square error (RMSE). The average R values of the PMH machine learning models for the three datasets are 0.824, 0.787, and 0.834, respectively, as shown in \autoref{table:Pearson-values-PDBbind}. These high R values validate the effectiveness and reliability of our PMH molecular representation. We also obtain low RMSE values (in units of kcal/mol), which compare the predicted binding energies with the experimental values. The binding energy is calculated from the given \( pK_d \) in the original data by multiplying it by a constant of 1.3633.

To enhance the predictive performance of our PMH machine learning models, we incorporate natural language processing (NLP)-based molecular features and develop an additional set of machine learning models. The pretrained NLP models generate molecular features using molecular sequences as input. Specifically, we utilize molecular features from transformer-based pretrained models for proteins \cite{rives2021biological} and small molecules \cite{winter2019learning}. These features are then integrated with the GBDT algorithm to create a new predictive model, referred to as model-seq. The modeling performance of this approach is presented in the third column of \autoref{table:Pearson-values-PDBbind}. The average R value of the PMH model exceeds that of the transformer-based machine learning model. Additionally, we create a consensus model by combining the strengths of the three models—model-PMH2, model-PMH5, and model-seq—by averaging their predictions to determine the final predicted binding affinity. The last column of \autoref{table:Pearson-values-PDBbind} shows the performance of the consensus model. The consensus model significantly boosts the performance of the PMH model, with an average R value of 0.832.

A series of advanced mathematical theories from algebraic topology and graph theory were employed to design molecular descriptors \cite{nguyen2019agl,meng2021persistent,liu2023persistent,cang2017topologynet}, leading to reliable machine learning models. Their success significantly relies on molecular characterization through topological invariants. Our machine learning model is comparable to these competitive models and demonstrates superior performance compared to a wide range of other published models. The Betti numbers from PMH include crucial topological invariants and provide additional mathematical analysis of molecular data. This significantly enhances the descriptive and predictive power of our molecular features.

\begin{table}[htb!]
	\small
	\centering
	\begin{tabular}{c| c c  c }
		\hline
		 \textbf{Dataset} &  \textbf{Total} & \textbf{Training set } & \textbf{Test set}\\
		 \hline
			PDBbind-v2007 \cite{cheng2009comparative} & 1300& 1105 &195\\
			PDBbind-v2013 \cite{li2014comparative}&  2959 &2764&195\\
			PDBbind-v2016 \cite{su2018comparative} &  4057 &3767&290 \\
			\hline
	\end{tabular}
\caption{Details of the datasets utilized for benckmark tests in this study.}
\label{table:datasets-information}
\end{table}

\begin{table}[htb!]
	\small
	\centering
	\begin{tabular}{c c c  c }
		\hline
		No. of estimators &  Max depth & Min. sample split & Learning rate\\
		20000& 7 &5 & 0.002\\
		\hline
	   Max features & Subsample size & Repetition &\\
		 Square root & 0.8&  20 times & \\
		\hline
	\end{tabular}
	\caption{Hyperparameters used for build gradient boosting regression models.}
	\label{table:GBDT-parameters}
\end{table}

\begin{table}[htb!]
	\small
	\centering
	\begin{tabular}{c | c c  c c c}
		\hline
		\textbf{Dataset} &  \textbf{PMH} & \textbf{Transformer} & \textbf{PMH+Transformer}\\
		\hline
		PDBbind-v2007 & 0.824(1.95)& 0.795(2.006) & \textbf{0.837(1.907)}\\
		PDBbind-v2013& 0.787(2.036)& 0.791(1.977)&  \textbf{0.807(1.982)}\\
		PDBbind-v2016 & 0.834(1.755)& 0.836(1.716)&  \textbf{0.851(1.701)}\\
		\hline
		Average & 0.815 (1.914) &0.807 (1.9) &   \textbf{0.832 (1.863)}\\
		\hline
	\end{tabular}
	\caption{Modeling performance of different strategies on the test sets of PDBbind-v2007, PDBbind-v2013 and PDBbind-v2016. Pearson correlation coefficient and root mean square error (unit, kcal/mol) are the two evaluation metrics.}
	\label{table:Pearson-values-PDBbind}
\end{table}

\begin{figure}[!htb]
	\centering
	\includegraphics[width=1\textwidth]{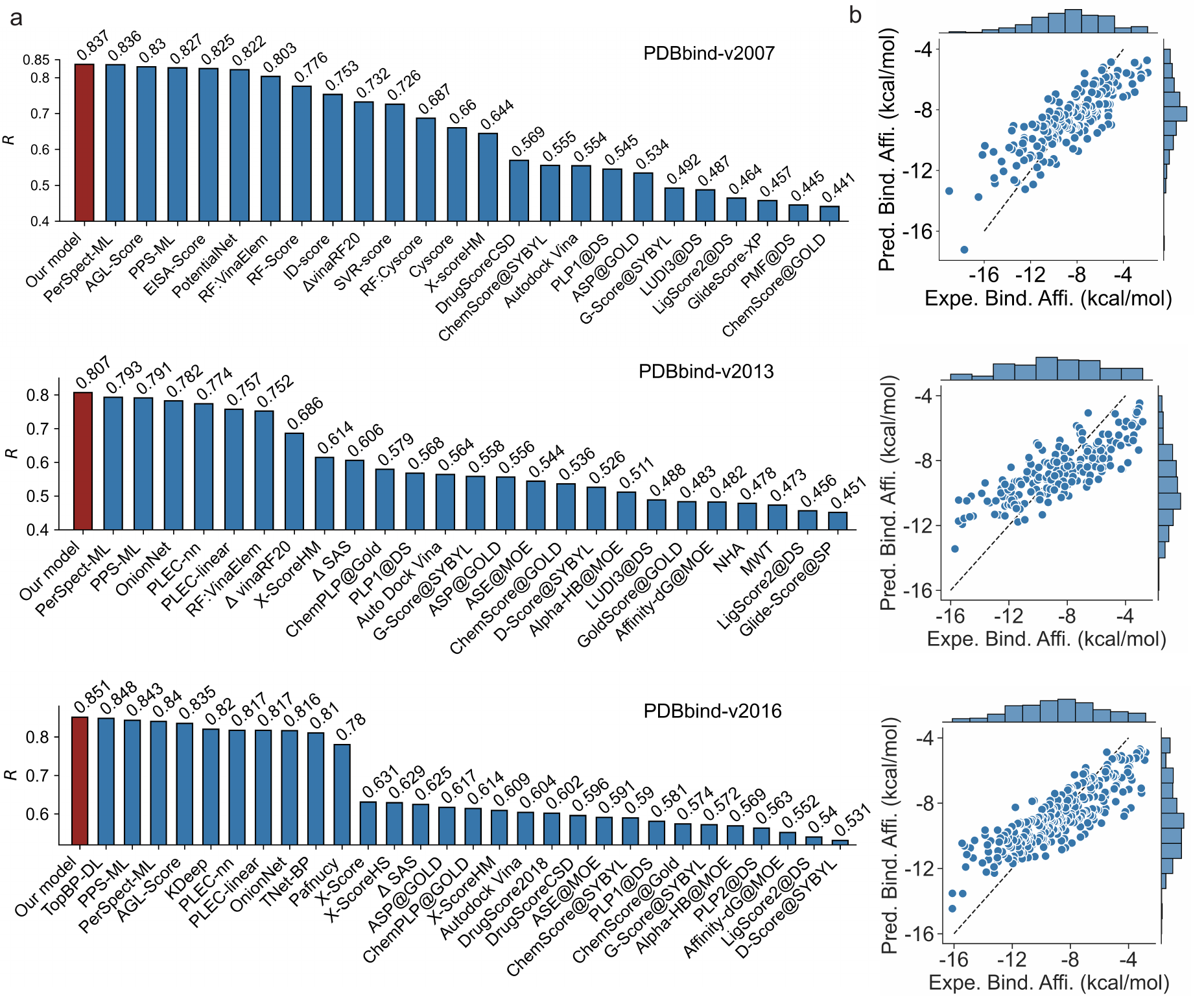}
	\caption{The prediction performance of my final machine learning model for three well-established protein-ligand binding affinity datasets including PDBbind-v2007, PDBbind-v2013, and PDBbind-v2016. The comparison of the experimental and predicted binding affinities for the three datasets are exhibited in the right column.
	}
	\label{Fig:R-comparison}
\end{figure}

We compare the performance of our consensus model with various models from the literature. \autoref{Fig:R-comparison} depicts these comparisons across the three PDBbind datasets. Our model outperforms a wide range of models and represents the state of the art. The second column in \autoref{Fig:R-comparison} shows the comparison between experimental energy and predictions from our final consensus model. The high consistency between the two sets of binding energies validates the accuracy and reliability of our machine learning model. Deep neural networks have advanced the development of the scientific community. Integrating our PMH molecular descriptors with deep neural networks has the potential to offer even more accurate predictive models.

\section{Discussion}\label{sec:Discussions}

Molecular feature extraction plays a fundamental role in AI-driven drug design. End-to-end molecular representation learning using deep neural networks, such as CNNs and GNNs \cite{wallach2015atomnet,ragoza2017protein,hassan2020rosenet}, has been developed and shown impressive capabilities in analyzing biological data and predicting protein-ligand binding affinities. In contrast, molecular featurization or engineering takes a different approach by encoding the structural, physical, chemical, or biological properties of molecules. Among the various molecular featurization methods, those based on advanced mathematics have emerged as a highly promising strategy for significantly enhancing molecular featurization capabilities.

Geometric, topological, and combinatorial invariants from differential geometry, algebraic topology, and algebraic graph theory have proven to be valuable mathematical techniques in this regard \cite{nguyen2019agl,nguyen2020review, meng2021persistent,liu2023persistent,cang2017topologynet}. These mathematical molecular representations are characterized by a higher level of abstraction and transferability. Persistent homology has demonstrated high effectiveness for molecular featurization by employing topological invariants and their persistence along a filtration process. The proposed persistent Mayer homology theory extends persistent homology while preserving its power. Specifically, the Betti number information derived from Mayer homology theory across different \(N\)-chain complexes includes the Betti number information of the usual simplicial homology when \(N\) is reduced to 2. The utilization of filtration induces multiscale representations by collecting sufficient Betti number information from persistent Mayer homology. The molecular features induced by PMH can further enhance the descriptive and predictive capacity of the original persistent homology theory and lead to learning models with boosted performance.

We primarily focus on PMH5 for simplicial complexes derived from Alpha complexes due to concerns about computation time and cost. To enhance efficiency, particularly when dealing with VR complexes, more efficient computer programs can be developed. Recently, a series of topological persistent Laplacian theories have been proposed \cite{meng2021persistent,liu2023persistent,wang2020persistent} and applied to the design of molecular features. The harmonic spectral information of these Laplacians recovers the Betti numbers of persistent homology, while the non-harmonic spectral information provides additional geometric insights into the data. Additionally, a Persistent Mayer Laplacian based on simplicial complexes was developed in our recent study \cite{shen2024persistent}. We will further explore the potential of Persistent Mayer Laplacian for molecular representation learning in our future work. Other significant advancements in molecular featurization include the application of persistent de Rham-Hodge Laplacians \cite{su2024persistent}. These methods demonstrate competitive performance and high potential. One key difference between these approaches and persistent homology or persistent Mayer homology is that they utilize curve-shaped data or continuum manifolds, whereas persistent homology and persistent Mayer homology focus on point cloud data. Our work in persistent Mayer homology advances the development of persistent homology and maintains its competitiveness.

\section{Methods}\label{sec:Methods}

\subsection{Persistent Mayer homology theory}

In this section, we introduce the mathematical theory of Persistent Mayer Homology (PMH). PMH builds on several key mathematical concepts, including $N$-chain complexes, Mayer homology, and Mayer Betti numbers. Persistent homology, a crucial theory in algebraic topology, serves as the foundation for PMH. However, PMH extends persistent homology by providing a broader mathematical framework that includes higher-order analysis.

\textbf{$N$-chain complex and generalized differential.}	
Let \( \mathbb{C} \) be the field of complex numbers, an $N$-chain complex is constructed as a graded \( \mathbb{C} \)-linear space \( C_* = (C_n)_{n \geq 0} \), equipped with linear maps \( d_n: C_n \rightarrow C_{n-1} \) of degree \(-1\). A key property of these differentials \( d_n \) is that when applied \( N \) times, they vanish, meaning \( d^N = 0 \). This property is fundamental to the structure of the $N$-chain complex and allows for the development of a homological theory based on these complexes.

Suppose \(K\) is a simplicial complex. The generalized differential of the simplicial complex is defined on the graded space \( C_* =(C_n)_{n\geq 0} \), where \( C_n \) is generated by the \( n \)-simplices of \( K \). The differential \( d_n \) acts on an \( n \)-simplex \( \langle v_0, v_1, \ldots, v_n \rangle \) by summing over the contributions of the \( (n-1) \)-simplices obtained by omitting one vertex at a time:

\[
d_n \langle v_0, v_1, \ldots, v_n \rangle = \sum_{i=0}^n \xi^i \langle v_0, \ldots, \hat{v}_i, \ldots, v_n \rangle,
\]
where \( \xi \) is the \( N \)-th primitive root of unity, and \( \hat{v}_i \) denotes the omission of the \( i \)-th vertex. This differential satisfies the property \( d^N = 0 \), a result stemming from the cyclic nature of the root of unity \( \xi \).

\textbf{Mayer Homology.} Given a simplicial complex \( K \), we can obtain the $N$-chain complex and generalized differential \( (C_*,d) \). For an integer \( q \) such that \( 1\leq q \leq N-1\), we have the following:
\[
C_{n+N-q}\xrightarrow{d^{N-q}}C_n\xrightarrow{d^q}C_{n-q}.
\]
Note that \( d^q\circ d^{N-q}=d^N=0\), the above is a classical chain complex. In general, we have
\[
\small
\begin{tikzcd}[column sep=large, row sep=large]
	\cdots \arrow[r, "d"] & C_{n+N-1} \arrow[r, "d^{N-1}"] \arrow[d, "d"] & C_n \arrow[r, "d"] \arrow[d, "id"] & C_{n-1} \arrow[r, "d^{N-1}"] \arrow[d, "d"] & C_{n-N} \arrow[r, "d"] \arrow[d, "id"] & C_{n-N-1} \arrow[d, "d"] \arrow[r] & \cdots \\
	\cdots \arrow[r, "d^2"] & C_{n+N-2} \arrow[r, "d^{N-2}"] \arrow[d, "d"] & C_n \arrow[r, "d^2"] \arrow[d, "id"] & C_{n-2} \arrow[r, "d^{N-2}"] \arrow[d, "d"] & C_{n-N} \arrow[r, "d^2"] \arrow[d, "id"] & C_{n-N-2} \arrow[d, "d"] \arrow[r] & \cdots \\
	\vdots & \vdots \arrow[d,"d"] & \vdots \arrow[d,"id"] & \vdots \arrow[d,"d"] & \vdots \arrow[d,"id"] & \vdots \arrow[d,"d"] & \vdots \\
	\cdots \arrow[r, "d^{N-2}"] & C_{n+2} \arrow[r, "d^2"] \arrow[d, "d"] & C_n \arrow[r, "d^{N-2}"] \arrow[d, "id"] & C_{n-N+2} \arrow[r, "d^2"] \arrow[d, "d"] & C_{n-N} \arrow[r, "d^{N-2}"] \arrow[d, "id"] & C_{n-2N+2} \arrow[d, "d"] \arrow[r] & \cdots \\
	\cdots \arrow[r, "d^{N-1}"] & C_{n+1} \arrow[r, "d"] & C_n \arrow[r, "d^{N-1}"] & C_{n-N+1} \arrow[r, "d"] & C_{n-N} \arrow[r, "d^{N-1}"] & C_{n-2N+1} \arrow[r] & \cdots . \\
\end{tikzcd}
\]
Thus we could give the \( (n,q) \)-th Mayer homology group of simplicial complex \( K \) :
\[
H_{n,q}(K) =H_{n,q}(C_*,d)=\frac{\ker d^q}{\mathrm{im}~d^{N-q}} =\frac{\{ x \in C_n \mid d^q x = 0 \}}{\{ d^{N-q} y \mid y \in C_{n+N-q} \}}.
\]

The rank of $H_{n,q}(K)$ is defined to be the Mayer Betti number, denoted as \( \beta_{n,q} \). 

Specifically, \( \beta_{n,q} \) is defined as the dimension of the vector space \( H_{n,q}(K) \):
\[
\beta_{n,q} = \dim H_{n,q}(K).
\]
These numbers provide a quantitative measure of the topological and geometric complexity of the complex in the context of Mayer homology.

\textbf{Persistent Mayer homology.} We will study the persistence of these topological features across different scales based on the filtration of simplicial complexes. For a non-decreasing real-valued function \( f: K \rightarrow \mathbb{R} \), we can obtain a filtration of simplicial complexes
\[
K_{a_1} \subseteq K_{a_2} \subseteq \cdots \subseteq K_{a_m} = K,
\]
for $a_{1}\leq a_{2}\leq \cdots\leq a_{m}$. Here,
\[
K_a = \{ \sigma \in K \mid f(\sigma) \leq a \}.
\]
As the filtration parameter\( a \) increases, the simplicial complex \( K_a \) grows, allowing for the analysis of how features of Mayer homology evolve.

Persistent Mayer homology is defined by examining the Mayer homology groups at different stages of the filtration. The persistent Mayer homology groups \( H_{n,q}^{a,b} \) measure the topological and geometric features that persist from the simplicial complex \( K_a \) to \( K_b \). Specifically, these groups are defined as:
\[
H_{n,q}^{a,b} = \text{im}\left(H_{n,q}(K_a) \rightarrow H_{n,q}(K_b)\right),\quad a\leq b.
\]
These persistent features can be visualized using persistence diagrams or barcodes, offering insights into the multi-scale topological and geometric structure of the data.

\subsection{Natural language processing (NLP) molecular descriptors}

We incorporate molecular descriptors derived from natural language processing (NLP) techniques to improve the performance of PMH machine learning models. A protein-ligand complex comprises the 3D structures of both the protein and the ligand. For these structures, sequence representations are utilized: amino acid sequences for the protein and SMILES strings for the ligand. NLP techniques are applied to design molecular features for each component. We then concatenate the molecular features from both parts into a single feature vector, which serves as the molecular descriptor for the entire protein-ligand complex.

\subsubsection{ESM transformer protein language model}

In recent years, significant progress has been made in modeling protein properties using large-scale protein transformer models trained on extensive datasets of protein sequences. A notable example is the ESM-2 transformer protein language model introduced by Rives et al. \cite{rives2021biological}, which extracts underlying biological properties and relationships by deeply analyzing vast quantities of amino acids protein sequences. This model was trained on a dataset consisting of 250 million sequences using a masked filling procedure and features a sophisticated architecture with 34 layers and an impressive 650 million parameters.

We utilized the ESM transformer model to generate sequence embeddings for the protein part. At each layer of the ESM model, a sequence of length \(L\) is encoded into a matrix of size \(1280 \times L\), excluding the start and terminal tokens. For our machine learning predictions, we used the sequence representation from the final (34th) layer and computed the average along the sequence length axis, resulting in a 1280-component vector.

\subsubsection{Transformer-based small molecular language model}

A transformer-based deep learning model was proposed to extract molecular representations \cite{chen2021extracting} and is a valuable tool for small molecule machine learning analysis \cite{hayes2024graph,shen2023svsbi}. This model was trained on over 700 million SMILES strings sourced from databases such as ChEMBL, PubChem, and ZINC. Three pretrained models were developed: model-C, model-CP, and model-CPZ. For each molecule, the model generates a \(256 \times 512\) dimensional matrix, where 256 represents the symbols associated with the molecule, and 512 denotes the vector dimension for each symbol. The molecular features are given by the average of the 256 vectors.

\section*{Funding Information}
This work was supported in part by NIH grants R01GM126189, R01AI164266, and R35GM148196, NSF grants DMS-2052983, DMS-1761320, DMS-2245903,  and IIS-1900473, NASA grant 80NSSC21M0023, MSU Foundation, Bristol-Myers Squibb 65109, and Pfizer.

\section*{Statement of Usage of Artificial Intelligence}

In the preparation of this manuscript, we utilized ChatGPT primarily to assist with improving the clarity and grammar of the English text. The AI tool was employed solely for language enhancement and did not contribute to the research content or data analysis.

\section*{Data Availability}

All data and the code needed to reproduce this paper's result can be found at\\ \href{https://github.com/WeilabMSU/PMH\_Bio}{https://github.com/WeilabMSU/PMH\_Bio}.

\section*{Author Contributions}

Hongsong Feng was responsible for the coding, computational experiments and writing. Li Shen was responsible for the coding and some writing. Jian Liu was responsible for the coding and some writing. Guo-Wei Wei was responsible for methodology, reviewing, editing, some writing and supervision.

\section*{Conflict of Interest}

The authors declare no conflict of interest.

\section*{ORCID}

\noindent Hongsong Feng- \url{https://orcid.org/0000-0001-8039-3059}

\noindent Li Shen - \url{https://orcid.org/0009-0003-0318-8328}

\noindent Jian Liu - \url{https://orcid.org/0000-0003-4362-5273}

\noindent  Guo-Wei Wei - \url{https://orcid.org/0000-0002-5781-2937}

\section*{Data and code availability}

All data and the code needed to reproduce this paper's result can be found at\\ \href{https://github.com/WeilabMSU/PMH\_Bio}{https://github.com/WeilabMSU/PMH\_Bio}.

\section*{Acknowledgment}	
This work was supported in part by NIH grants R01GM126189, R01AI164266, and R35GM148196, NSF grants DMS-2052983, DMS-1761320, DMS-2245903,  and IIS-1900473, NASA grant 80NSSC21M0023, MSU Foundation, Bristol-Myers Squibb 65109, and Pfizer.

 \vspace{1cm}
	

\end{document}